\documentclass[aps,pre,preprint,showpacs,amssymb,floatfix]{revtex4}
\usepackage{amsmath,graphics,epsfig}

\newcommand{\half}{\frac{1}{2}}

\newcommand{\gae}{\hbox{\lower0.7ex\hbox{$\sim$}\llap{\raise0.4ex\hbox{$>$}}}}
\newcommand{\lae}{\hbox{\lower0.7ex\hbox{$\sim$}\llap{\raise0.4ex\hbox{$<$}}}}
\begin{document}
\title{Specific heat of the simple-cubic Ising model}
\author{Xiaomei Feng$^{1,2}$ and Henk W.J. Bl\"ote$^{1,3}$}
\affiliation{$^1$Faculty of Applied Sciences, Delft University of
Technology,\\  P.O. Box 5046, 2600 GA Delft, The Netherlands}
\affiliation{$^2$Nanjing University of Aeronautics and Astronautics,
\\College of Materials Science and Technology, 29 Yudao St.,
210016 Nanjing, P.R. China}
\affiliation{$^3$Lorentz Institute, Leiden University,
  P.O. Box 9506, 2300 RA Leiden, The Netherlands}             
\email{henk@lorentz.leidenuniv.nl}
\date{\today} 
\begin{abstract}
We provide an expression quantitatively describing the specific heat
of the Ising model on the simple-cubic lattice in the critical region. 
This expression is based on finite-size scaling of numerical results
obtained by means of a Monte Carlo method. It agrees satisfactorily
with series expansions and with a set of experimental results. Our
results include a determination of the universal amplitude ratio of
the specific-heat divergences at both sides of the critical point.

\end{abstract}
\maketitle 

\section{introduction}
\label{Intro}
Though real magnetic systems were supposed to be Heisenberg-like, the
Ising model was originally introduced \cite{I} as a simplified model
of magnetic ordering, because its relative simplicity offers 
better possibilities for a theoretical analysis.  In later years,
it was found, however, that Ising-like magnetic systems do exist.
This is because real systems consist of spins embedded in a crystal
lattice, and the resulting anisotropy field due to the neighboring
charges may lift the O(3) symmetry of an unperturbed spin. 
Depending on the character of the perturbation, the spin may have an
`easy axis' or an `easy plane'. Here we consider the former case,
which leads to Ising-like behavior.

In many cases, the perturbation is relatively small and the system
will approximately behave Heisenberg-like, except near an ordering
transition where the paramagnetic state transforms into a long-range 
ordered one. Near the transition, crossover \cite{ST,PF,FW} occurs to
Ising-like behavior. The critical singularities are then described
by the Ising set of critical exponents.
In some other cases, the perturbation due to the crystal field is 
so strong that the magnetic spins assume a true Ising character.
This situation occurs when the ionic angular momentum $\vec{S}$ is
described by a spin quantum number $S>\half$, and the crystal field
lifts the degeneracy of the $S_z$ eigenstates such that the $S_z=\pm S$
doublet is lowest in energy, with the higher levels so far away that
they play no role, even in the presence of exchange interactions 
between neighboring spins. Then the low-lying doublet can be described
by an effective spin-1/2 Ising Hamiltonian. This situation is known to 
occur for the Co$^{2+}$ ion in a tetrahedral coordination. It occurs also
for some rare-earth ions like Dy$^{3+}$ and Yb$^{3+}$ in a sufficiently
strong crystal field, with the provision that here the magnetic moments
are due to spin as well as orbital angular momentum, and should thus be
denoted $\vec{J}$ instead of $\vec{S}$.

If such ions are embedded in a crystal structure for which theoretical
predictions for the thermodynamical properties such as the specific
heat exist, comparison with experiments may be possible \cite{dJM,WPW}.
Such comparisons were made for dysprosium phosphate \cite{JCW,WPW} 
and for some alkali cobalt halides \cite{WBRH,BH}. These systems
were found to behave, at least approximately, as the Ising models on
the diamond lattice and the simple-cubic lattice respectively.

The best way to obtain theoretical results for the thermodynamic properties
of these models would obviously be an exact solution, but this is known
to be a very difficult problem. It is thus noteworthy that it was claimed 
recently by Zhang \cite{Zhang} that a conjectured exact solution was found
for the three-dimensional Ising model. However, Perk \cite{JP} and Wu et 
al.~\cite{Wuea} pointed out that Zhang's result for the free energy and 
the underlying arguments are flawed. Here it may be added that Zhang's
result for the critical point of the simple-cubic Ising models is not
compatible with independent and mutually consistent numerical estimates
\cite{BC3,DB}. The difference with Zhang's result exceeds the estimated
numerical accuracies \cite{BC3,DB} by several orders of magnitude.

In the absence of an exact solution, one may still resort
to approximations. At temperatures sufficiently far above and below the
critical point, excellent approximations exist in the form of series
expansion of the partition function or the free energy, such as given in
Refs.~\onlinecite{BCGS} and \onlinecite{AF} for the model on the 
simple-cubic lattice. In the critical region, the series of a finite
length become inaccurate, and a method to extrapolate these series on 
the basis of a critical scaling assumption, such as used by Butera and
Comi \cite{BC}, is needed. In the case of
Rb$_3$CoCl$_5$ (rubidium cobalt chloride) \cite{BH} the required
theoretical prediction for the specific heat near criticality was also
obtained this way. A similar analysis has been performed for the
specific heat of DyPO$_4$ (dysprosium phosphate) \cite{JCW,WPW}, which
was instead compared with series expansions for the diamond lattice.
However, these specific-heat analyses were conducted at a time that the
value of the critical exponent $\alpha$ was not well known, for instance,
$\alpha$ was set to zero in Ref.~\onlinecite{BH}. 
Moreover, Wegner's correction to scaling \cite{FW2} was not included.

In order to obtain accurate predictions for the heat capacity in the
critical region, one may apply Monte Carlo simulations. Cluster
simulation methods \cite{SW,W}, which strongly reduce critical slowing
down, allow statistically accurate simulations in the critical region.
Extrapolation of the finite-size simulation data to the thermodynamic
limit is possible if the simulations cover a range of finite sizes
exceeding the correlation length. Whereas this still excludes, as a
result of the divergence of the correlation length, a narrow temperature
range about the critical point, one may attempt to describe the
extrapolated data by means of a scaling formula. The present work
reports our efforts along this line for the case of the energy and the
specific heat of the Ising model on the simple-cubic lattice.

In Sec.~\ref{secnum} we describe our Monte Carlo simulations, and the
extrapolation to infinite system size. The derivation of scaling formulas
for the energy and the specific heat, and the data analysis in terms of 
these formulas, are presented in Sec.~\ref{secsca}. Section \ref{secdis}
discusses the numerical accuracies, provides comparisons with results
from series expansions and with a set of experimental results, and ends
with a few concluding remarks.

\section{Numerical technique}
\label{secnum}
The reduced Hamiltonian (Hamiltonian divided by $kT$) of the Ising
model is denoted
\begin{equation}
{\mathcal H}(K)= -K \sum_{\langle i,j\rangle } s_i s_j
\label{H0}
\end{equation}
where the indices $i$ and $j$ label nearest-neighbor lattice sites
on the simple-cubic lattice.
The sum is on all nearest-neighbor pairs, and the spins $s_k$ can
assume values $\pm 1$. The coupling is defined by $K\equiv J/kT$
where $J$ is minus the energy of a pair of parallel nearest-neighbor
spins, $k$ the Boltzmann constant, and $T$ the temperature.
The canonical reduced free energy density $f$ is equal to
\begin{equation}
f = \frac{1}{N} \ln Z\, ,~~~~ Z = \sum_{\{ S \} } e^{-{\mathcal H}(K)}
\label{fr}
\end{equation}
where $Z$ is the partition function, $N$ the number of spins, and 
the sum is on all spin configurations $\{ S \}$. The energy $E$ and
the specific heat $C$ per particle, as expressed in dimensionless units,
follow from the derivatives of $f$ to $K$:
\begin{equation}
\frac{E}{J} \,=\frac{E}{kTK}  =- \frac{d f}{d K}\, ,~~~~ 
\frac{C}{k} \,= K^{2} \frac{d^2 f}{d K^2} \, .
\label{ec}
\end{equation}
\subsection{Monte Carlo calculations}
\label{secmc}
Substitution of Eqs.~(\ref{fr}) and (\ref{H0}) in Eqs.~(\ref{ec})
leads to
\begin{equation}
\frac{E}{J}= \frac{1}{NK} \langle {\mathcal H} \rangle
\label{ec1}
\end{equation}
and
\begin{equation}
\frac{C}{k} = \frac{1}{N}
(\langle {\mathcal H}^2 \rangle - \langle {\mathcal H} \rangle^2) \, ,
\label{ec2}
\end{equation}
where the ensemble averages $\langle x \rangle$, which are defined as
\begin{equation}
\langle x \rangle \equiv \frac{1}{Z} \sum_{\{S\}}x e^{-{\mathcal H}(K)}\,,
\end{equation}
can be sampled directly using importance sampling.

The simulations involved the sampling of the energy, as well
as its square, for $L \times L \times L$ Ising systems on 
simple-cubic lattices, with periodic boundary conditions. 
The system sizes were chosen as powers of 2 in the range
$ 4 \leq L \leq 128$, and in addition as $L=6$ and 12. About $10^7$
samples were taken for $L \leq 16$, $2 \times 10^6$ for $L=32$,
$3 \times 10^5$ for $L=64$, and $5 \times 10^4$ for $L=128$.
Each sample was preceded by a number or Wolff cluster steps and/or
Metropolis sweeps, depending on the value of $K$ in comparison with
the critical coupling $K_{\rm c}\approx 0.2216546$ \cite{DB}. For
$K<< K_{\rm c}$,  Wolff clusters tend to be very small and only
Metropolis sweeps were applied, and for $K>K_{\rm c}$ only Wolff
cluster steps. In the intermediate range, a few Metropolis sweeps
were supplemented with a number of Wolff cluster steps.
The number of Wolff clusters was chosen roughly equal to the
inverse of the relative Wolff cluster size. The coupling $K$ was 
given some 50 different values chosen to cover a wide range about 
the critical point.

\subsection{Extrapolation}
\label{secext}
The analysis of the  numerical finite-size data was done on the 
basis of well-documented finite-size scaling methods \cite{FSS}.
For non-critical systems with sizes $L$ exceeding the correlation
length, the data for the energy should approximately behave as
\begin{equation}
E(K,L)=E(K,\infty) + a(K) e^{-L/\xi(K)} + \cdots
\end{equation}
from which the extrapolated energy $E(K,\infty)$ was obtained by
means of a least-squares analysis. A small-system-size cutoff was
applied when necessary to obtain a satisfactory residual $\chi^2$.
This cutoff varied between $L=6$ far away from the critical point, 
and $L=32$ at a distance $|K-K_{\rm c}| \approx 0.005$ from
the  critical point. No reliable extrapolations were obtained for
$|K-K_{\rm c}|$ less than a few times $10^{-3}$, with the exception
of $K=K_{\rm c}$,
where one expects that the finite-size-dependent energy converges
as a power of $L$, which again enables extrapolation to $L=\infty$.
Typical estimated accuracies of the extrapolated results for $E/J$ 
are in the order of $10^{-5}$.

The same extrapolation procedure was applied to the
finite-size data for the specific heat with $|K-K_{\rm c}| > 0.005$. 
Typical accuracies of the extrapolated results for $C/k$ are estimated
as at most a few times $10^{-4}$ for $K<0.2$ and $K>0.25$, up to a few
times $10^{-3}$ in the vicinity $K_{\rm c}$.
The extrapolated data are listed in the Appendix. 

\section{Scaling and least-squares analysis}
\label{secsca}
\subsection{Derivation from renormalization theory}
The analysis of the  extrapolated data was done on the
basis of scaling as derived from renormalization theory.
The relevant equations follow from the assumptions that the picture
described in the following paragraph is valid.

The free energy density $f(T_1,T_2,\cdots)$ of the infinite system,
expressed as a function of thermodynamic parameters $T_j$ 
($j=1,~2,~\cdots$), can be written as the sum of an analytic
part $f_a(T_1,T_2,\cdots)$ and a singular part $f_s$. The
singular part can be written $f_s(t_1,t_2,\cdots)$ as a function
of Wegner's \cite{FW1} nonlinear scaling fields $t_j$, which are
analytic functions of the $T_j$ in a neighborhood of a critical point
under investigation. Thus
\begin{equation}
f(T_1,T_2,\cdots)=f_a(T_1,T_2,\cdots)+f_s(t_1,t_2,\cdots)
\end{equation}
The singular part satisfies the scaling equation as implied by the
renormalization theory. A rescaling of the linear dimensions by a factor
$b$ thus leads to
\begin{equation}
f_s(t_1,t_2,\cdots)=b^{-d}f_s(b^{y_1}t_1,b^{y_2}t_2,\cdots)
\end{equation}
where $d$ is the dimensionality and the $y_j$ are the renormalization
exponents associated with the scaling fields $t_j$, with the temperature
exponent $y_1$ positive, and the other exponents negative.
The choice $b=|t_1|^{-1/y_1}$ thus yields
\begin{equation}
f_s(t_1,t_2,\cdots)=|t_1|^{d/y_1}f_s(\pm 1,|t_1|^{-y_2/y_1}t_2,\cdots)
\label{scfs}
\end{equation}
where $\pm 1$ has the sign of $t_1$.
Furthermore, $f_s(x_1,x_2,x_3,\cdots)$ is an analytic function in a
neighborhood of the $x_1=1$, $x_2=0$,  $x_3=0$, $\cdots$.

On the basis of this set of assumptions, we may Taylor expand the free
energy in powers of the arguments $T_j$ and $t_j$, and then expand the
$t_j$'s in the $T_j$'s, resulting in an expression depending only on 
the physical temperature fields, but with expansion coefficients that
remain to be determined. We follow this procedure, restricting number 
of scaling fields in the expansion of Eq.~(\ref{scfs}) to two,
namely the temperature field $t\equiv t_1$ and the irrelevant field
$\tilde{u} \equiv t_2$. The corresponding exponents are denoted $y_t$ and
$y_u$ respectively. The temperature exponent $y_t$ determines the leading
singularity in the temperature-induced ordering transition, while the
irrelevant exponent $y_u$ generates Wegner's correction to scaling
\cite{FW2}. Expansion of the right-hand side of Eq.~(\ref{scfs}) 
thus yields
\begin{equation}
f_s(t,\tilde{u})=|t|^{d/y_t} \sum_{j} (j!)^{-1}        
f_s^{0,j}(\pm 1,0) |t|^{-jy_u/y_t}\tilde{u}^j \, ,
\label{fsex}
\end{equation}
where $f_s^{0,j}$ is the $j$th derivative of $f_s$ to its second
argument. The scaling fields are expanded as analytic power series
in the temperature-like parameter $t_0$, defined by
\begin{equation}
t_0 \equiv \Delta K/K\, , ~~~~ \Delta K \equiv K-K_{\rm c} \, .
\label{t0def}
\end{equation}
The analytic part of the free energy $f_a$ can be expanded directly
in powers of $\Delta K$. The resulting expansion of the total free 
energy density in powers of $\Delta K$ and $t$ can be expressed in $K$,
the only variable physical temperature parameter in our problem,
as given by the Hamiltonian (\ref{H0}). Differentiation
of the resulting expansion of the free energy density to $K$ yields
the dimensionless energy $E/J$. For $d=3$ dimensions, the leading terms
are included in
\begin{displaymath}
-E(K,\infty)/J=
\hspace{110mm}
\end{displaymath}
\begin{equation}
\sum_{j=0,1,\cdots}e_j(\Delta K)^j+\, \frac{d|t|}{d K}\,
 a_{\pm} |t|^{(3-y_t)/y_t}+ b_{\pm} u |t|^{(3-y_t-y_u)/y_t} +
 p_{\pm}  u^2 |t|^{(3-y_t-2y_u)/y_t} + \cdots \, .
\label{eexp}
\end{equation}
where we have included the first three terms in the sum on $j$ in 
Eq.~(\ref{fsex}), and $u$ is an analytic function of $t_0$ related to
$\tilde{u}$ by
\begin{equation}
\frac{d-y_u}{y_t} \, f^{0,1}_{s}(\pm 1,0) \, \frac{d|t|}{dK} \,
\tilde{u}= b_{\pm}u
\label{uut}
\end{equation}
The dimensionless specific heat $C/k$ of the model (\ref{H0}) satisfies
\begin{equation}
\frac{C(K,\infty)}{k}=K^2 \,\frac{d^2f(K,\infty)}{dK^2}=-\frac{K^2}{J}
\,\frac{dE}{dK}
\end{equation}
and its expansion thus follows by differentiation of the energy,
Eq.~(\ref{eexp}). This leads to
\begin{displaymath}
\frac{C(K,\infty)}{kK^2}= \sum_{j=1,2,\cdots} j e_j (\Delta K)^{j-1}+ 
\frac{3-y_t}{y_t} \, \left( \frac{d|t|}{d K} \right)^2 \,
 a_{\pm} |t|^{(3-2y_t)/y_t} \, + \, \frac{d^2|t|}{d K^2} \,
a_{\pm}|t|^{(3-y_t)/y_t} \, + 
\hspace{20mm}
\end{displaymath}
\begin{equation}
\hspace*{20mm}
\frac{3-y_t-y_u}{y_t} \, \frac{d|t|}{d K} \, u b_{\pm} |t|^{(3-2y_t-y_u)/y_t} 
+ \frac{du}{d K} \, b_{\pm} |t|^{(3-y_t-y_u)/y_t}  + \cdots \, ,
\label{sphex}
\end{equation}
The parameters $t$ and $u$, and their derivatives as they appear in 
Eqs.~(\ref{eexp}) and (\ref{sphex}), are expanded in powers of $t_0$ as
\begin{displaymath}
t=\sum_{j=1,2,\cdots} w_j t_0^j\, ,~~~~
\frac{d|t|}{d K}=\pm\frac{K_{\rm c}}{K^2}\sum_{j=1,2,\cdots}jw_jt_0^{j-1}\,,
\end{displaymath}
\begin{equation}
\frac{d^2|t|}{d K^2}=
\pm\frac{K_{\rm c}}{K^2}\sum_{j=2,3,\cdots}j(j-1)w_jt_0^{j-2}
 \mp\frac{K_{\rm c}}{K^3}\sum_{j=1,2,\cdots}jw_jt_0^{j-1}\,,
\label{tsum}
\end{equation}
where $\pm$ stands for the sign of $t$, $\mp$ for its opposite, and
\begin{equation}
 u=\sum_{j=0,1,\cdots} u_j t_0^j\, , ~~~~
\frac{d u}{d K} =\frac{K_{\rm c}}{K^2} \sum_{j=1,2,\cdots}ju_j t_0^{j-1}\, .
\label{usum}
\end{equation}
The scales of $t$ and $u$ are determined by setting $w_1=u_0= 1$.

\subsection{Fits}
\label{secfit}
Whereas Eq.~(\ref{fsex}) includes, in principle, infinitely many terms,
for numerical work it is necessary to truncate the expansion of $f_s$,
as well as those of $f_a$ and the scaling fields, at a finite order.
Expression (\ref{eexp}) for the energy already contains the implicit
simplification that there is only one irrelevant field, and that
the expansion of $f_s(\pm 1,x)$ is truncated at second order.
Moreover, higher orders in the expansion of the temperature
derivative of the irrelevant field were neglected. We shall reconsider
these simplifications in Sec.~\ref{errdis}. No further simplifications
were made in the derivation of Eq.~(\ref{sphex}) from Eq.~(\ref{eexp}).

Many attempts were made to fit
Eqs.~(\ref{eexp}) and (\ref{sphex}) to the numerical data, using different
ranges of $K$, and different sets of parameters as determined by the
orders at which the expansions were truncated. The unknown parameters
in each set were determined by means of a Levenberg-Marquardt nonlinear
least-squares analysis. Since Eqs.~(\ref{eexp}) and (\ref{sphex}) depend 
on the same parameters, the data for the energy and the heat capacity were
simultaneously fitted by one set of parameters.  

A fit was considered satisfactory if it met three criteria: first,
the residual $\chi^2$ has to be compatible with the number of degrees
of freedom; second, there should be sufficiently large ranges of overlap
with the accurate predictions from the low- and high-temperature series
expansions; and third, at least the amplitudes of the leading terms in
the fit formulas should be reasonably stable under variations of the
$K$-interval and of the number of correction terms in the temperature
field and the analytic background.
In Table \ref{tab:parms} we list the smallest satisfactory set of
parameters thus obtained. We skipped the ellipses in Eqs.~(\ref{eexp})
and (\ref{sphex}), and included terms up to order $j=4$ in the expansion
of $t$, up to $j=2$ in that of $u$, and up to $j=5$ in the analytic
parts expressed by the first sums in Eqs.~(\ref{eexp}) and (\ref{sphex}).
The residual of this fit was $\chi^2=53.5$, to be compared with the
number of degrees of freedom $d_{\rm f}=84$. Since possible correlations
between specific heat and energy data could influence the estimation
of the errors in the fitted parameters, we have analyzed the
correlations between the deviations of the energy and of the specific
heat with respect to the fit formula. We find a correlation coefficient
of -0.066 which is not significant, and does not provide a reason to
reconsider our error estimates.

\begin{table}[ht]
\caption{Values of the parameters in the fit according to
Eqs.~(\ref{eexp}) and (\ref{sphex})
to Monte Carlo data in the interval $0.15 \leq K \leq 0.60$. The error
estimates given in the last column are not only based on statistics,
but also on the variations of the parameter values due to changes of the
fit interval and the number of parameters. In two cases the estimated
error exceeds the parameter value and no error is quoted. While these
values have no physical meaning, they are still useful for the evaluation
of the specific heat and the energy. The values of $y_t$, $y_u$, and
$K_{\rm c}$ were taken from Ref.~\onlinecite{DB}.}
\label{tab:parms}
\begin{center}
\begin{tabular}{||c|c|c||}
\hline
parameter         &       value              & error margin \\
\hline
$w_1$             & $    1     $             &   fixed      \\
$w_2$             & $    0.662300$           &   0.06       \\
$w_3$             & $    0.160415$           &   0.09       \\
$w_4$             & $    0.008397$           &   ------     \\
$u_0$             & $    1       $           &   fixed      \\
$u_1$             & $   -2.673700$           &   0.6        \\
$a_-$             & $    1.466642$           &   0.016      \\
$a_+$             & $    2.758572$           &   0.012      \\
$b_-$             & $    0.923100$           &   0.2        \\
$b_+$             & $   -2.694381$           &   0.4        \\
$p_-$             & $   -1.440041$           &   0.4        \\
$p_+$             & $   -2.345305$           &   0.8        \\
$e_0$             & $    0.990604$           &   0.000004   \\
$e_1$             & $  -27.847250$           &   0.8        \\
$e_2$             & $  110.506127$           &   12         \\
$e_3$             & $ -193.032628$           &   50         \\
$e_4$             & $  186.624090$           &   100        \\
$e_5$             & $  -80.141986$           &   ------     \\
$y_t$             & $    1.587   $           &   fixed      \\
$y_u$             & $   -0.82    $           &   fixed      \\
$K_{\rm c}$       & $   0.2216546$           &   fixed      \\
\hline
\end{tabular}
\end{center}
\end{table}
During the least-squares analysis, we found that some parameter 
values changed significantly when the $K$-interval and/or the numbers
of parameters in the expansions of $t$ and of the analytic background
were varied. Such shifts were sometimes comparable to the error margins 
as estimated from statistics based on the accuracy of the Monte Carlo
results. This applied in particular to those of the $w_j$ and the $e_j$
with $j>2$. In this respect the amplitudes $a_+$, $a_-$, $e_0$, $e_1$
and, to some extent, $b_+$ and $b_-$ were better behaved.
The error estimates listed in Table \ref{tab:parms} take into 
account the variation of the parameter values between these fits.

%

\section{Discussion}
\label{secdis}
\subsection{Choice of parameters and their error margins}
\label{errdis}
Equation (\ref{fsex}) and the fits of $E$ and $C$ use only one irrelevant
field, while, according to Newman and Riedel \cite{NR}, corrections to
scaling could also arise from a second irrelevant field $u'$ with exponent
$y_{u'} \approx 2 y_u$. We note that corrections generated in first order
of $u'$ would thus, in the present context, be practically indistinguishable
from those generated in second order by $u$. For this reason, we have not
included a separate term containing $u'$.
Furthermore, the energy, Eq.~(\ref{eexp}), neglects a contribution due
to the possible $K$-dependence of the irrelevant field. Such a term
behaves as $|t|^{(3-y_u)/y_t}$ and is thus a factor $|t|$ smaller than
the leading correction. The third-order correction in $u$, which is
also neglected, has nearly the same exponent.

Another correction that was neglected is one with an integer exponent
$y''=-2$, associated with the discreteness of the cubic lattice. The
presence of such corrections could modify the higher-order correction
amplitudes given in Table \ref{tab:parms}, but the $\chi^2$ criterion
did not yield indications that a term with $y''=-2$ should be included.

Some insight in the relative importance of the corrections due to
different orders of the irrelevant field can be obtained by 
comparing the fit including the second order of $u$, as given in
Table \ref{tab:parms}, to fits including up to the first order.
Reasonable fits, as following from the $\chi^2$ criterion, could only
be be obtained by including three more coefficients $e_j$ or $w_j$.
Moreover, these coefficients tended to assume much larger values.
For this reason, we prefer the fit up to second order in $u$, although
the fit up to first order also yields a satisfactory numerical
representation of the critical energy and specific heat.

Only the parameters $a_-$, $a_+$, $c_0$ and $c_1$, describing the leading
few orders of $E$ and $C$, were about the same for both types of fits.
It is thus clear that not too much physical significance should be given
to the subleading and higher-order
parameters given in Table \ref{tab:parms}, except that they provide a
numerical description of $E$ and $C$ in the critical region.

The relative errors in the amplitudes $p_-$ and $p_+$ of the second 
order term in $u$, as given in Table \ref{tab:parms}, are appreciable,
and far exceed those of the first-order amplitudes $b_-$ and $b_+$.
For this reason we believe that it is not necessary to include a
third-order correction, or other terms with approximately the same
exponent.

\subsection{Comparison with existing results}
\subsubsection{Series expansions }
\begin{figure}[ht]
\includegraphics[width=10cm]{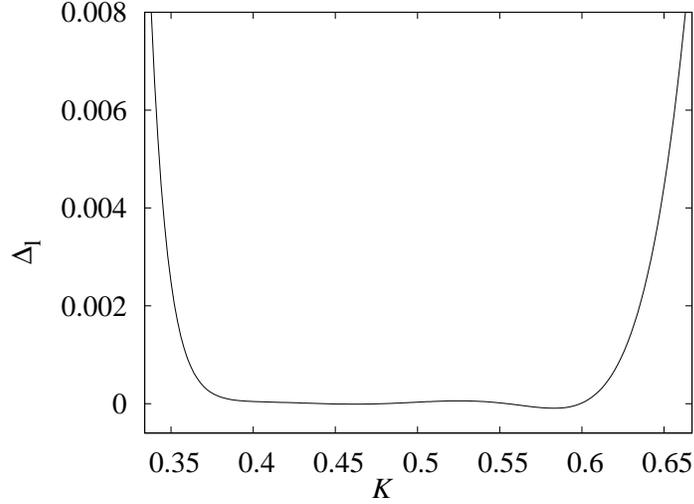}
\caption{Difference $\Delta_{\rm l} \equiv (C_{\rm LTE} -C_{\rm fit})/k$
between the specific heat of the Ising model as obtained from the
low-temperature series of the energy and from the present least-squares
analysis according to Eq.~(\ref{sphex}). The difference $\Delta_{\rm l}$
is at most $10^{-4}$ in the interval $0.39<K<0.60$.}
\label{comlt}
\end{figure}

Numerical evaluation of Eq.~(\ref{sphex}) allows comparison with
results from series expansions. The low-temperature series
for the energy is provided by Bhanot et al.~\cite{BCGS} up to order
25 in $e^{-2K}$. The specific heat, as obtained by differentiation
of this series, is in good agreement with Eq.~(\ref{sphex}) in the
interval $0.39<K<0.60$. The differences, which are shown in
Fig.~\ref{comlt}, do not exceed $10^{-4}$.
For  $K>0.60$,  outside the range of the least-squares fit, our
representation of the specific heat with Eq.~(\ref{sphex})
is no longer accurate and the differences increase sharply.
The increasing differences for $K<0.30$ are due to the truncation of
the low-temperature series to 25 terms.

\begin{figure}[ht]
\includegraphics[width=10cm]{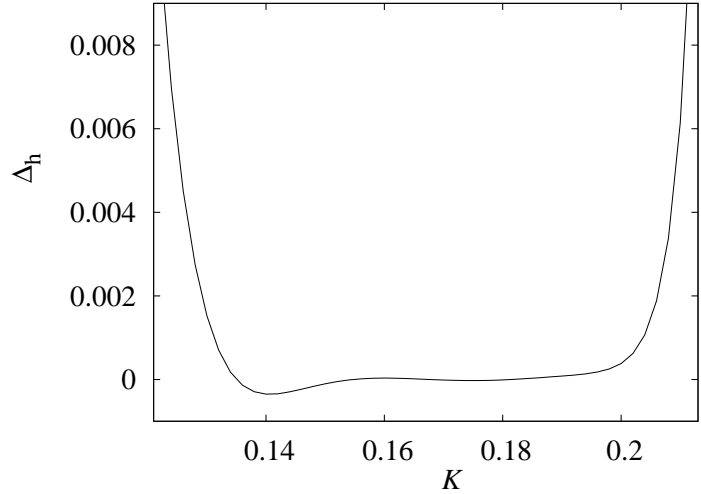}
\caption{Difference $\Delta_{\rm h} \equiv (C_{\rm HTE} -C_{\rm fit})/k$
between the specific heat of the Ising model as obtained from the
high-temperature series of the energy and from the present least-squares
analysis according to Eq.~(\ref{sphex}). The difference $\Delta_{\rm h}$
is at most $10^{-4}$ in the interval $0.15<K<0.19$.}
\label{comht}
\end{figure}

For temperatures above the critical point, a comparison can be made
based on the series expansion up to order 46 of the free energy
as provided by Arisue and Fujiwara \cite{AF}, with the help of
Eq.~(\ref{ec}). The differences with Eq.~(\ref{sphex}) are less than
$10^{-4}$ interval  $0.15<K<0.19$,  as plotted in Fig.~\ref{comht}.
For $K<0.15$, outside the range of the fit, Eq.~(\ref{sphex}) rapidly
loses its accuracy. The increasing differences for $K>0.19$ are due
to the truncation of the series.

\subsubsection{Amplitude ratios and analytic background}
The fit up to first order in $u$ yielded a universal amplitude ratio
$a_-/a_+=0.540~(5)$, which is to be compared to the result of the fit
including the second order of $u$, which is $a_-/a_+=0.532~(7)$ as
follows from the parameter values in Table \ref{tab:parms}.
Based on the consistency between these two results, we believe that
the latter result $a_-/a_+=0.532~(7)$ is reliable.
This result is close to an estimate 0.541~(14) by Bagnuls et al.~\cite{ft}
from field theory, and to the result 0.523~(9) obtained by Liu and
Fisher \cite{LiuF} based on series expansions, and slightly smaller
than 0.560~(10) as determined from Monte Carlo data by Hasenbusch and
Pinn \cite{HP}.

Another universal ratio that can be constructed from the results in
Table \ref{tab:parms} concerns the corrections-to-scaling amplitudes.
The data in the table suggest $b_-/b_+=-0.34~(9)$, which differs
considerably from $b_-/b_+=-0.96~(25)$ as obtained by Bagnuls et
al.~\cite{ft} (note the sign difference with respect to the notation
used by Bagnuls et al., which is related to the factor $d|t|/dK$ in our
Eq.~(\ref{sphex})). The sign of this amplitude ratio is in agreement
with the conclusions of Liu and Fisher \cite{LiuF2}.

As noted in Sec.~\ref{errdis}, there may be corrections to scaling 
governed by an irrelevant field $u'$ with exponent $y_{u'} \approx 2 y_u$,
and thus indistinguishable from contributions in second order of $u$.
It is thus possible that the amplitudes $p_+$ and $p_-$ as given in
Table \ref{tab:parms} contain contributions due to the field $u'$.
Therefore, the resulting ratio $p_-/p_+=0.61~(24)$ may not qualify as
a universal amplitude ratio.

Our result for the critical energy, $e_0=0.990604~(4)$, can be compared
with results obtained from series analysis. It is slightly smaller than
the result $e_0=0.99218~(15)$ obtained by Sykes et al.~\cite{SHMH},
slightly larger than $e_0=0.9902~(1)$ found by Liu and Fisher \cite{LiuF},
and in agreement with $e_0=0.991~(1)$ found by  Butera and Comi \cite{BC2}.
Our result is also consistent with the Monte Carlo estimates $e_0=0.990~(4)$
due to Jensen and Mouritsen \cite{JM}, and $e_0=0.9904~(8)$ due to
Hasenbusch and Pinn \cite{HP}.

\subsubsection{Comparison with experimental results for Rb$_3$CoCl$_5$ }
\begin{figure}[ht]
\hspace*{-20mm} \includegraphics[width=20cm]{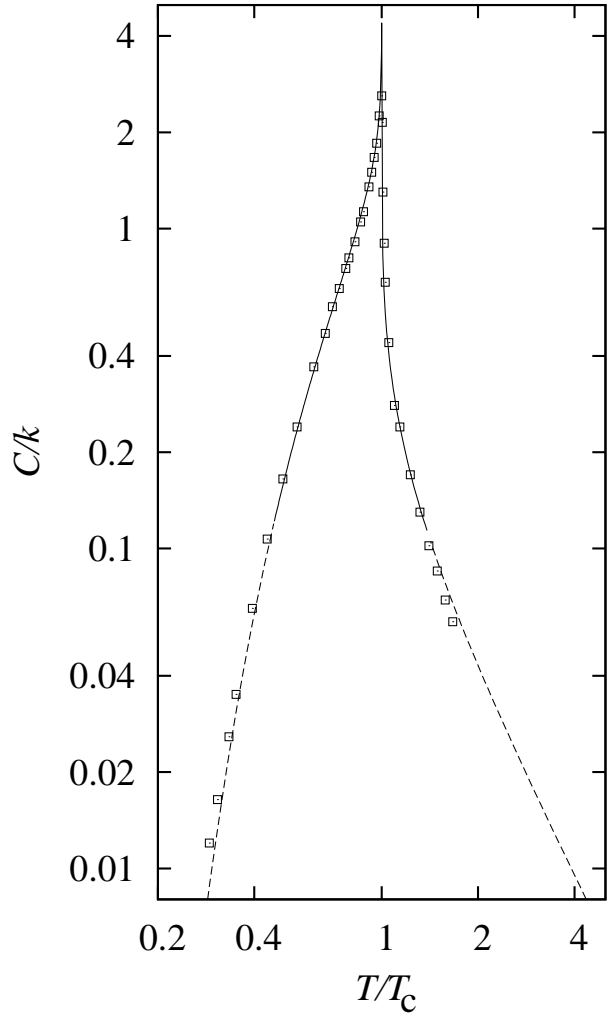}
\caption{Specific heat of the Ising model on the simple-cubic lattice.
Logarithmic scales are used because of the large variation of the 
specific heat with temperature. The data points are existing 
experimental results \cite{BH} for Rb$_3$CoCl$_5$. The full line
represents the scaling form Eq.~(\ref{sphex}) with the parameters
defined in Table \ref{tab:parms}. The dashed lines at the lower 
left and right are obtained from low- and high-temperature series
expansions {\protect\cite{BCGS,AF}} of the 
the free energy.}
\label{comex}
\end{figure}

As implied in the Introduction, the magnetic Co$^{2+}$ ions in
rubidium cobalt chloride assume a spin-1/2 Ising character.
This has been experimentally confirmed \cite{BBSZ} in the related
compound Cs$_3$CoCl$_5$. The magnetic moments are aligned along the $c$
direction of the tetragonal crystal structure.
The Co$^{2+}$ ions are arranged in a simple Bravais lattice, with
equivalent positions \cite{Pow}. Furthermore, electron-spin resonance
results \cite{SHHB} for Cs$_3$CoCl$_5$ showed that the exchange
interaction with the two nearest neighbors in the crystallographic $c$
direction has the same magnitude as that with the four nearest neighbors
in the $aa$ plane, so that one may expect that the theoretical results
for the simple-cubic Ising model are applicable. Specific-heat and
magnetic susceptibility measurements \cite{BH} on Rb$_3$CoCl$_5$
showed that a phase transition to an antiferromagnetic phase occurs
at $T_{\rm c}=1.14$ K. It was indeed found that the specific heat 
(which does not depend on the sign of $K$) did agree with the
theoretical predictions available at that time.
These predictions were based on series expansions due to Baker \cite{JB}
and Sykes \cite{MS}, and on the assumption that the specific-heat exponent
$\alpha=0$. In view of later results for the specific-heat exponent, as
well as the effect of Wegner's correction \cite{FW2}, the comparison made
in Ref.~\onlinecite{BH} may thus not be considered as entirely
satisfactory. In Fig.~\ref{comex} we show the experimental data together
with Eq.~(\ref{sphex}), as well as results from the low- and 
high-temperature series. This comparison with the experimental data,
which involves only one adjustable parameter, the critical temperature,
shows that the specific heat of Rb$_3$CoCl$_5$ 
agrees reasonably well with the predictions for the simple-cubic
Ising model. The data in Fig.~\ref{comex} suggest small deviations at
low as well as at high temperatures, but there the specific heat becomes
very small, so that the experimental error margins, which include the 
uncertainty of heat capacity of the empty apparatus, become appreciable.
A comparison of the experimental data listed in Ref.~\onlinecite{BH}
with the results from Eq.~(\ref{sphex}) show that deviations up to a
few percent occur also in the range $0.9<K/K_{\rm c}<1.2$. But these
deviations do not display an obvious systematical trend, and may
possibly be attributed to the fact that the measured heat capacity
is, near criticality, the result of integration of a highly nonlinear
function over a nonzero temperature range.

It thus seems that new experiments on Rb$_3$CoCl$_5$ are needed to
firmly establish deviations with respect to the predictions for the
simple-cubic Ising model. Such deviations would be a logical consequence
of the tetragonal symmetry of Rb$_3$CoCl$_5$, which implies that there
is no reason why the coupling in the $c$-direction should be precisely
equal to that in the $a$ direction.  Also the presence of interactions
with further neighbor spins, which include small magnetic dipole-dipole
interactions, should lead to deviations.

\subsection{Conclusion}
The formula Eq.~(\ref{sphex}), supplemented
by Eqs.~(\ref{tsum}), (\ref{usum}),  and (\ref{t0def})
and by the parameter values in Table \ref{tab:parms}, describes the 
specific heat of the three-dimensional Ising model in the interval
$0.15 < J/kT < 0.60$. Comparisons with low- and high-temperature
series expansions yield satisfactory agreement in the intervals
$0.15 < J/kT < 0.19$ and $0.39 < J/kT < 0.60$ respectively. The 
differences between Eq.~(\ref{sphex}) and the results from series
expansions are at most $10^{-4}$ in the mentioned intervals. These
differences are smaller than the statistical errors in the Monte Carlo
results on which Table \ref{tab:parms} is based, as may be expected
since the number of 100 data points far exceeds the number of 16 free
parameters in the fit formula, so that in effect averaging occurs.
Since Eq.~(\ref{sphex}) continues to satisfactorily describe the
extrapolated Monte Carlo data until a distance
$|K-K_{\rm c}| \approx 0.005$ from the critical point,
we conclude that the error margin in Eq.~(\ref{sphex}) does not
exceed that of the Monte Carlo data, i.e., it will be limited to
at most a few times  $10^{-3}$ at $|K-K_{\rm c}| > 0.005$. Larger
uncertainties are expected for $|K-K_{\rm c}| <0.005$ because
of the error margins in the critical amplitudes, exponents and
temperature. Taking into account these numerical uncertainties,
Eq.~(\ref{sphex}) can be used in the interval $0.15<J/kT < 0.60$ for
comparison with experiments on systems that are described by the
simple-cubic Ising Hamiltonian.

In addition, our results show that Monte Carlo simulations can be used
to determine the universal leading amplitude ratios $a_-/a_+$ and even
the nonasymptotic ratio $b_-/b_+$. Thus far, the correction amplitudes
have been studied by means of series analysis, field theory, and
crossover scaling, see e.g., Refs.~\onlinecite{LiuF2,ft,SD,AKST}.
\acknowledgments
HB thanks J.M.J. van Leeuwen for valuable discussions, and the Lorentz
Fund (The Netherlands) for financial support.


\appendix
\section*{Appendix: Extrapolated energy and specific heat}
\begin{table}[ht]
\caption{Extrapolated values of the dimensionless energy density of the 
infinite, simple-cubic Ising model as a function of the coupling $K$.
Estimated error bounds are included.}
\label{tab:ener}
\vspace{-2mm}
\begin{center}
\begin{tabular}{||c|c|c||c|c|c||}
\hline
  $K$       & $-E(K)$   & error    & 
  $K$       & $-E(K)$   & error    \\
\hline
  0.12      & 0.382236  & 0.000020 & 
  0.13      & 0.419100  & 0.000109 \\
  0.14      & 0.457696  & 0.000007 & 
  0.15      & 0.498271  & 0.000098 \\
  0.16      & 0.541261  & 0.000008 & 
  0.166     & 0.568535  & 0.000448 \\
  0.17      & 0.587433  & 0.000007 & 
  0.172     & 0.597108  & 0.000047 \\
  0.178     & 0.627251  & 0.000049 & 
  0.18      & 0.637719  & 0.000009 \\
  0.184     & 0.659300  & 0.000052 & 
  0.19      & 0.693640  & 0.000008 \\
  0.195     & 0.724504  & 0.000007 & 
  0.196     & 0.730987  & 0.000060 \\
  0.2       & 0.757945  & 0.000005 & 
  0.202     & 0.77224   & 0.000011 \\
  0.205     & 0.794807  & 0.000016 & 
  0.21      & 0.836533  & 0.000053 \\
  0.213     & 0.865007  & 0.002836 & 
  0.22165460& 0.990604  & 0.000004 \\
  0.224     & 1.135886  & 0.000083 & 
  0.225     & 1.184077  & 0.000113 \\
  0.226     & 1.228851  & 0.002835 & 
  0.227     & 1.270902  & 0.001567 \\
  0.228     & 1.310761  & 0.000069 & 
  0.229     & 1.348806  & 0.000084 \\
  0.23      & 1.385036  & 0.000010 & 
  0.232     & 1.453408  & 0.000128 \\
  0.235     & 1.547095  & 0.000082 & 
  0.24      & 1.684411  & 0.000006 \\
  0.25      & 1.908377  & 0.000006 & 
  0.26      & 2.084422  & 0.000011 \\
  0.268     & 2.200165  & 0.000117 & 
  0.27      & 2.226207  & 0.000025 \\
  0.28      & 2.342331  & 0.000013 & 
  0.29      & 2.438392  & 0.000020 \\
  0.3       & 2.518570  & 0.000013 & 
  0.31      & 2.585908  & 0.000017 \\
  0.32      & 2.643039  & 0.000137 & 
  0.33      & 2.691193  & 0.000121 \\
  0.34      & 2.732442  & 0.000779 & 
  0.35      & 2.767639  & 0.000008 \\
  0.36      & 2.797886  & 0.000224 & 
  0.37      & 2.823926  & 0.000119 \\
  0.38      & 2.846340  & 0.000055 & 
  0.39      & 2.865798  & 0.000070 \\
  0.4       & 2.882622  & 0.000005 & 
  0.42      & 2.909917  & 0.000030 \\
  0.44      & 2.930623  & 0.000034 & 
  0.45      & 2.939050  & 0.000003 \\
  0.46      & 2.946380  & 0.000022 & 
  0.48      & 2.958427  & 0.000035 \\
  0.5       & 2.967777  & 0.000002 & 
  0.6       & 2.990703  & 0.000001 \\
  0.65      & 2.994958  & 0.000001 & 
  0.7       & 2.997255  & 0.000001 \\
\hline
\end{tabular}
\end{center}
\end{table}

\begin{table}[ht]
\caption{Extrapolated values of the dimensionless specific heat of the
infinite, simple-cubic Ising model as a function of the coupling $K$.
Estimated error bounds are included.}
\label{tab:spht}
\begin{center}
\begin{tabular}{||c|c|c||c|c|c||}
\hline
  $K$       & $C(K)/k$ & error    & 
  $K$       & $C(K)/k$ & error    \\
\hline
  0.12      & 0.05212  &  0.00004 & 
  0.14      & 0.07736  &  0.00005 \\
  0.15      & 0.09382  &  0.00013 & 
  0.16      & 0.11381  &  0.00004 \\
  0.166     & 0.12812  &  0.00008 & 
  0.17      & 0.13882  &  0.00005 \\
  0.172     & 0.14458  &  0.00010 & 
  0.178     & 0.16391  &  0.00012 \\
  0.18      & 0.17094  &  0.00012 & 
  0.19      & 0.21445  &  0.00035 \\
  0.195     & 0.24361  &  0.00031 & 
  0.2       & 0.27950  &  0.00110 \\
  0.205     & 0.32850  &  0.00093 & 
  0.208     & 0.36500  &  0.00100 \\
  0.21      & 0.39573  &  0.00114 & 
  0.214     & 0.48216  &  0.00165 \\
  0.215     & 0.51335  &  0.00261 & 
  0.216     & 0.54767  &  0.00157 \\
  0.217     & 0.58500  &  0.00200 & 
  0.227     & 2.11009  &  0.00675 \\
  0.228     & 2.01738  &  0.00323 & 
  0.229     & 1.95000  &  0.01000 \\
  0.23      & 1.87829  &  0.00300 & 
  0.232     & 1.76800  &  0.00150 \\
  0.235     & 1.64500  &  0.01000 & 
  0.238     & 1.52800  &  0.00300 \\
  0.24      & 1.46600  &  0.00200 & 
  0.25      & 1.23138  &  0.00041 \\
  0.26      & 1.06294  &  0.00031 & 
  0.27      & 0.93268  &  0.00200 \\
  0.28      & 0.82520  &  0.00060 & 
  0.29      & 0.73597  &  0.00094 \\
  0.3       & 0.66021  &  0.00072 & 
  0.31      & 0.59383  &  0.00085 \\
  0.35      & 0.39923  &  0.00039 & 
  0.36      & 0.36325  &  0.00036 \\
  0.37      & 0.33045  &  0.00024 & 
  0.38      & 0.30110  &  0.00020 \\
  0.39      & 0.27460  &  0.00020 & 
  0.4       & 0.25058  &  0.00025 \\
  0.42      & 0.20890  &  0.00020 & 
  0.44      & 0.17428  &  0.00013 \\
  0.45      & 0.15910  &  0.00020 & 
  0.46      & 0.14567  &  0.00012 \\
  0.48      & 0.12158  &  0.00010 & 
  0.5       & 0.10160  &  0.00010 \\
  0.55      & 0.06470  &  0.00015 & 
  0.6       & 0.04109  &  0.00010 \\
  0.65      & 0.02593  &  0.00008 & 
  0.7       & 0.01632  &  0.00004 \\
\hline
\end{tabular}
\end{center}
\end{table}


\begin{thebibliography} {99}
\bibitem{I}
E. Ising, Z. Physik {\bf 31}, 253 (1925).
\bibitem{ST}
M. Suzuki, Phys. Lett. A {\bf 35}, 23 (1971);
M. Suzuki and F. Tanaka, Prog. Theor. Phys. {\bf 3}, 1085 (1973).
\bibitem{PF}
M. E. Fisher and P. Pfeuty, Phys. Rev. B {\bf 6}, 1889 (1972).
\bibitem{FW}
F. J. Wegner, Phys. Rev. B {\bf 6}, 1891 (1972).
\bibitem{dJM}
L. J. de Jongh and A. R. Miedema, Adv. Phys. {\bf 23}, 1 (1974).
\bibitem{WPW}
W. P. Wolf, Braz. J. of Phys. [online] {\bf 30}, n. 4,
794 (2000).
\bibitem{JCW}
J. C. Wright, H. W. Moos, J. H. Colwell, B. W. Magnum and D. D.
Thornton, Phys. Rev. B {\bf 3}, 843 (1971).
\bibitem{WBRH} 
R. F. Wielinga, H. W. J. Bl\"ote, J. A. Roest and W. J. Huiskamp,
Physica (Amsterdam) {\bf 34} 223 (1967).
\bibitem{BH} 
H. W. J. Bl\"ote and W. J. Huiskamp,
Phys. Lett. A {\bf 29}, 304 (1969).
\bibitem{Zhang}
Z.-D. Zhang, Phil. Mag. {\bf 87}, 5309 (2008).
\bibitem{JP} 
J. H. H. Perk, arXiv:0811.1802.v2 (2008).
\bibitem{Wuea}
F. Y. Wu, B. M. McCoy, M. E. Fisher and L. Chayes,
Phil. Mag. {\bf 88}, 3093 (2008).
\bibitem{BC3}
P. Butera and M. Comi, Phys. Rev. B {\bf 62}, 14837 (2004).
\bibitem{DB}
Y. Deng and H. W. J. Bl\"ote, Phys. Rev. E {\bf 68}, 036125 (2003),
and references therein.
\bibitem{BCGS}
G. Bhanot, M. Creutz and J. Lacki, Phys. Rev. Lett. {\bf 69}, 1841 (1992);
G. Bhanot, M. Creutz, U. Glassner and K. Schilling, 
Phys. Rev. B {\bf 49}, 12909 (1994).
\bibitem{AF}
H. Arisue and T. Fujiwara, Phys. Rev. E {\bf 67}, 066109 (2003).
There is a typo in the 42th order term; the correct value appears in
hep-lat/0209002 (2002).
\bibitem{BC}
P. Butera and M. Comi, Phys. Rev. B {\bf 69}, 174416 (2004).
\bibitem{FW2}
F. J. Wegner, Phys. Rev. B {\bf 5}, 4529 (1972).
\bibitem{SW}
R. H. Swendsen and J. S. Wang, Phys. Rev. Lett. {\bf 58}, 86 (1987).
\bibitem{W}
U. Wolff, Phys. Rev. Lett. {\bf 62}, 361 (1989).
\bibitem{FSS}
For reviews, see e.g. M. P. Nightingale in {\it Finite-Size Scaling and
Numerical Simulation of Statistical Systems}, ed. V. Privman (World
Scientific, Singapore 1990),
and M. N. Barber in {\it Phase Transitions and Critical Phenomena},
eds. C. Domb and J. L. Lebowitz (Academic, New York 1983), Vol.~8.
\bibitem{FW1}
F. J. Wegner, in {\em Phase Transitions and Critical Phenomena},
eds. C. Domb and M. S. Green (Academic, N.Y. 1976), Vol. 6.
\bibitem{NR}
K. E. Newman and E. K. Riedel, Phys. Rev. B {\ bf 30}, 6615 (1984). 
\bibitem{ft}
C. Bagnuls, C. Bervillier, D. I. Meiron and B. G. Nickel, Phys. Rev.
B {\bf 35}, 3585 (1987).
\bibitem{LiuF}
A. J. Liu and M. E. Fisher, Physica A (Amsterdam) {\bf 156}, 35 (1989).
\bibitem{HP}
M. Hasenbusch and A. Pinn, J. Phys. A {\bf 31}, 6157 (1998).
\bibitem{LiuF2}
A. Liu and M. E. Fisher, J. Stat. Phys. {\bf 58}, 431 (1990).
\bibitem{SHMH}
M. F. Sykes, D. L. Hunter,  D. S. McKenzie and B. R. Heap, 
J. Phys. A {\bf 5}, 667 (1972).
\bibitem{BC2}
P. Butera and M. Comi, Phys. Rev. B {\bf 60}, 6749 (1999).
\bibitem{JM}
S. J. K. Jensen and O. G. Mouritsen, J. Phys. A {\bf 15}, 2631 (1982).
\bibitem{BBSZ}
H. G. Beljers, P. F. Bongers, R. P. van Stapele and H. Zijlstra,
Phys. Letters {\bf 12}, 81 (1964).
\bibitem{Pow}
H. M. Powell and A. F. Wells, J. Chem. Soc. 359 (1935).
\bibitem{SHHB}
R. P. van Stapele, J. C. M. Henning, G. E. G. Hardeman and  P. F. Bongers,
Phys. Rev. {\bf 150}, 310 (1966).
\bibitem{JB}
J. M. Baker. Phys. Rev. {\bf 129}, 99 (1963).
\bibitem{MS}
M. F. Sykes, private communication cited in Ref.~\onlinecite{BH}.
\bibitem{SD}
R. Schloms and V. Dohm, Phys. Rev. B {\bf 42}, 6142 (1990).
\bibitem{AKST}
M. A. Anisimov, S. B. Kiselev, J. V. Sengers and S. Tang,
Physica A (Amsterdam) {\bf 188}, 487 (1992).
\end{thebibliography}
\end{document}